\documentclass[12pt, a4paper]{article}

\usepackage[margin=2.5cm]{geometry}
\usepackage{times}           
\usepackage{setspace}
\usepackage{amsmath, amssymb}
\usepackage{graphicx}
\usepackage{booktabs}        
\usepackage{tabularx}
\usepackage{multirow}
\usepackage{array}
\usepackage{caption}
\usepackage{subcaption}
\usepackage{float}
\usepackage{hyperref}
\hypersetup{
    colorlinks = true,
    linkcolor  = blue,
    citecolor  = blue,
    urlcolor   = blue
}
\usepackage{natbib}         
\usepackage{xcolor}

\title{\textbf{Graph-Based Multi-Omics Integration Improves Subtype Recovery and
Survival Prediction Over Classical Integration Strategies in TCGA-BRCA}}

\author{
  Taha Ahmad\textsuperscript{1}\\[2pt]
  \small\textsuperscript{1}Middle East Technical University
}
\date{}

\begin{document}
\maketitle

\newpage
\begin{abstract}

\textbf{Background.}
Breast cancer comprises at least five molecular subtypes with distinct
prognoses, yet PAM50 classification relies on transcriptomics alone. Whether
integrating DNA methylation and copy number data improves subtype recovery and
survival prediction over single-omic baselines remains an open question.

\textbf{Methods.}
We applied Similarity Network Fusion (SNF) to $n = 644$ TCGA-BRCA patients
with matched RNA-seq, 450k DNA methylation, and GISTIC2 copy number profiles.
Per-modality patient similarity networks were iteratively fused ($K = 20$,
$T = 20$, $\mu = 0.5$) and partitioned by spectral clustering; $k = 2$ was
pre-specified on eigengap and silhouette criteria. SNF was benchmarked against
RNA-only, CNV-only, methylation-only, and early concatenation baselines using
PAM50 NMI for subtype recovery and out-of-fold concordance index (OOF
C-index) from a Ridge Cox model with $N = 1{,}000$ bootstrap CIs for
pairwise comparisons.

\textbf{Results.}
SNF produced a stable two-cluster partition (stability ARI $= 1.00$, silhouette
$= 0.228$), with NMI $= 0.495$ versus PAM50, exceeding RNA-only ($0.428$) and
early concatenation ($0.175$). IHC receptor data confirmed cluster biology
independently (ER$^{+}$: $92.8\%$ vs $15.6\%$; triple-negative: $1.0\%$ vs
$45.4\%$; both $p < 10^{-100}$). SNF achieved an OOF C-index of $0.681$
(95\% CI $0.610$--$0.760$), significantly outperforming CNV-only
($\Delta = +0.122$, CI $0.020$--$0.211$); the advantage over RNA-only
($\Delta = +0.049$, CI $-0.036$--$0.144$) did not exclude zero.

\textbf{Conclusion.}
Graph-based multi-omics fusion recovers breast cancer subtype biology more
faithfully than feature concatenation and outperforms the weakest unimodal
baselines in survival prediction. The improvement over RNA-seq alone is
positive in direction but not yet statistically conclusive at this cohort size,
pointing to the trade-off between integration complexity and the sample sizes
needed to quantify its marginal benefit.


\end{abstract}

\newpage
\section{Introduction}

Breast cancer is among the most clinically and molecularly heterogeneous
solid tumours. Gene expression studies conducted in the early 2000s
established that what pathologists had long grouped together under one
diagnosis is better understood as a collection of at least five distinct
molecular diseases that is, Luminal~A, Luminal~B, HER2-enriched, Basal-like, and
Normal-like with each having its own characteristic prognosis and sensitivity
to therapy \citep{Perou2000, Sorlie2001}. This classification, standardised
later into the PAM50 gene signature \citep{Parker2009}, is now routinely
used in clinical practice, yet it rests on transcriptomic data alone. The
question of whether integrating additional molecular layers can sharpen that
picture or more practically, improve survival prediction has motivated
a decade of multi-omics work on the TCGA-BRCA cohort
\citep{TCGANetwork2012}.

The argument for using more than one omic modality is straightforward.
Transcriptomics captures what genes are being expressed, but a tumour's
behaviour is also shaped by which genomic regions have been amplified or
deleted and by whether gene promoters are silenced through methylation.
Copy number alterations (CNAs) drive focal amplifications of oncogenes
such as \textit{ERBB2} and deletions of tumour suppressors, effects that
may not be fully visible in expression data. DNA methylation at CpG sites
provides a more stable, epigenetic record of cell lineage and can mark
genes for silencing even when transcript levels appear normal. Each
modality therefore adds information the others do not fully contain, and
there is a reasonable prior expectation that combining them should do
better than any one alone.

The practical challenge is how to combine them. The simplest approach
is feature concatenation that is stacking all three feature matrices
column-wise and reducing dimensionality before clustering. This is
fast and easy to implement, but it can be dominated by the noisiest or
highest-dimensional modality, and it discards the within-modality
structure that makes similarity-based methods effective. At the other end
of the complexity spectrum, generative models such as iCluster
\citep{Shen2009} and factor-based methods such as MOFA
impose a joint latent structure across modalities but
require specifying a parametric form and are computationally demanding
at scale. Similarity Network Fusion (SNF), proposed by Wang and colleagues
\citep{Wang2014}, occupies a practical middle ground: each modality is
first converted into a patient similarity network independently, and then
the networks are fused iteratively through a message-passing procedure
that lets each network update and correct the others. The procedure is
non-parametric, modality-agnostic, and runs in $O(n^{2})$ time, making it
tractable for cohorts of several hundred patients.

SNF was originally demonstrated on three datasets including a glioma
cohort, where it outperformed single-omic clustering by a substantial
margin \citep{Wang2014}. Subsequent work has applied it to several cancer
types using the TCGA pan-cancer resource, generally finding that the fused
network recovers known subtypes better than any single modality. What the
literature lacks, however, is a careful quantitative comparison that
simultaneously (i) uses a well-characterised external ground truth (PAM50),
(ii) assesses survival prediction with cross-validated concordance indices
and bootstrap confidence intervals for pairwise differences, and (iii)
benchmarks SNF against both single-omic and early-integration baselines in
a single reproducible pipeline. Published comparisons often report point
estimates without uncertainty, making it difficult to judge whether
observed gains are statistically distinguishable from zero.

This study addresses those gaps using 644 TCGA-BRCA patients with matched
RNA-seq, 450k DNA methylation, and GISTIC2 copy number profiles. We ask
three questions. First, does SNF recover PAM50 subtypes better than
single-omic or concatenation-based clustering? Second, does fusing all
three modalities improve out-of-sample survival prediction relative to
the best single modality, and is any advantage statistically distinguishable
from zero at this cohort size? Third, how sensitive are the results to
SNF hyper-parameters (neighbourhood size $K$, number of diffusion steps
$T$, and feature count)? We report all comparisons with bootstrap
confidence intervals and treat the RNA-seq baseline as the primary
benchmark, given the established role of transcriptomics in breast cancer
subtyping. Our central hypothesis is that iterative graph fusion by
preserving within-modality structure and propagating cross-modality
agreement will recover molecular subtypes more faithfully and improve
survival discrimination relative to both single-omic and feature-concatenation
baselines, with measurable effect sizes and quantified uncertainty.

\newpage
\section{Materials and Methods}

\subsection{Dataset and cohort assembly}

Raw data were downloaded from the NCI Genomic Data Commons (GDC) using the
GDC Data Transfer Tool (v1.6). We obtained three data types for
TCGA-BRCA: RNA-seq HTSeq read counts (Workflow: HTSeq - Counts, GRCh38),
Illumina 450k DNA methylation $\beta$-values (Workflow: SeSAMe), and GISTIC2
somatic copy number segment files. PAM50 intrinsic subtype labels were
retrieved from the TCGA PanCanAtlas supplemental tables
\citep{Hoadley2018}. The three molecular matrices were intersected to retain
patients present in all three modalities, then further restricted to patients
with complete overall survival (OS) data and a minimum follow-up of 30~days,
giving a final cohort of $n = 644$ patients. Of these, $n = 568$ (88.2\%)
had PAM50 labels available from the PanCanAtlas resource; the remaining
76 patients (11.8\%) lacked PAM50 assignment (consistent with sample
quality thresholds in the original TCGA processing) and were excluded
from ARI and NMI calculations but retained in all survival analyses.
The derivation of this cohort
is recorded in \texttt{data/processed/final\_cohort.txt}; all downstream
analyses operate on this fixed patient list.

\subsection{Preprocessing}

\paragraph{RNA-seq.}
Raw counts were normalised to counts per million (CPM) and
$\log_{2}(\text{CPM}+1)$-transformed. Genes with CPM $< 1$ in more than
80\% of samples were removed as low-expression noise. The remaining genes
were ranked by median absolute deviation (MAD) across patients, and the top
$5{,}000$ were retained. Values were then $z$-scored per gene.

\paragraph{DNA methylation.}
$\beta$-values were filtered using the Illumina 450k probe annotation from
the Zhou laboratory (hg38) \citep{Zhou2017}. Removed probe classes were:
(i) probes on sex chromosomes (chrX, chrY); (ii) probes overlapping common
SNPs (Zhou MASK\_snp5\_common or MASK\_general flags); (iii) rs* control
probes. Probes with more than 20\% missing values were also discarded.
Batch effects across TCGA plates were corrected with ComBat as implemented
in the Python package \texttt{neurocombat-sklearn}. The top $5{,}000$ CpG
probes by MAD were selected and $z$-scored per probe.

\paragraph{Copy number.}
GISTIC2 segment-level copy number values ($\log_2$ ratio) were loaded
directly from the per-sample segment files. Features were filtered to the
top $5{,}000$ genomic segments by MAD and $z$-scored. No additional
normalisation was applied because GISTIC2 output is already
library-size-corrected.

\paragraph{Pre-fusion winsorisation.}
Before constructing similarity matrices, each modality's $z$-scored matrix
was winsorised at $\pm 5\sigma$ to prevent focal copy number amplifications
from driving cosine distances and dominating the affinity kernel.

\subsection{Similarity network construction and fusion}

For each of the three modalities, a $644 \times 644$ patient similarity
matrix was computed using the bounded exponential kernel of Wang~et~al.
\citep{Wang2014}:
\[
W(i,j) = \exp\!\left(-\frac{d(i,j)^{2}}{\mu \cdot \bar{\varepsilon}(i,j)}\right),
\]
where $d(i,j)$ is the Euclidean distance between patients $i$ and $j$,
$\bar{\varepsilon}(i,j)$ is the mean distance from $i$ and $j$ to their
respective $K$ nearest neighbours (local scaling), and $\mu = 0.5$ is a
fixed hyperparameter. The full-rank matrix $W$ was converted to a sparse
$K$-NN graph $P$ by retaining only the $K = 20$ nearest-neighbour affinities
and normalising rows to sum to one.

The three sparse graphs $P_{\text{RNA}}$, $P_{\text{CNV}}$, and
$P_{\text{Meth}}$ were fused using the iterative message-passing procedure of
SNF implemented in the Python package \texttt{snfpy} \citep{Ross2019}:
\[
P_l^{(t+1)} = P_l^{(t)} \cdot \frac{1}{m-1}\left(\sum_{k \neq l} W_k\right) \cdot \left(P_l^{(t)}\right)^\top,
\quad l = 1, \ldots, m,
\]
run for $T = 20$ diffusion iterations with $K = 20$. The final fused
affinity matrix $W_{\text{fused}}$ (644\,$\times$\,644) was the average of
the three graphs after convergence. All SNF hyperparameters ($K$, $T$, $\mu$)
were fixed at the defaults from the original paper and were not tuned on
outcome data.

\subsection{Spectral embedding and clustering}

The normalised graph Laplacian of $W_{\text{fused}}$ was eigendecomposed
and the top 50 eigenvectors were retained as a 644\,$\times$\,50 spectral
embedding. The same procedure was applied to each single-modality affinity
matrix to produce baseline embeddings.

Spectral clustering was applied to the fused embedding for $k = 2$--$6$
partitions. The primary partition number $k = 2$ was pre-specified based on
two independent criteria: (i) a clear eigengap at position 2 in the
eigenvalue spectrum of $W_{\text{fused}}$, and (ii) the highest average
silhouette coefficient ($s = 0.228$ at $k = 2$; all other $k$ values were
below 0.06). Although PAM50 defines five intrinsic subtypes, the dominant
axis of molecular variance in TCGA-BRCA is the Luminal versus Basal-like
contrast; $k = 2$ therefore captures the primary biological axis while
remaining statistically stable at $n = 644$. The study does not claim to
resolve the full five-subtype taxonomy, but rather to evaluate how faithfully
the dominant axis is recovered across integration strategies. Cluster stability was measured with adjusted Rand index (ARI) across 100
sub-sampled replicates (80\% of patients drawn without replacement); the
$k = 2$ solution was perfectly stable (mean ARI$= 1.00$, SD$= 0.00$).
Subtype concordance with PAM50 labels was measured by ARI and normalised
mutual information (NMI), defined as follows.

NMI between cluster labels $U$ and reference labels $V$ is computed as
\[
\text{NMI}(U,V) = \frac{2\,I(U;\,V)}{H(U)+H(V)},
\]
where $I(U;V) = H(U)+H(V)-H(U,V)$ is the mutual information and $H(\cdot)$
is Shannon entropy; NMI $\in [0,1]$ with 1 denoting perfect agreement.

The adjusted Rand index corrects the raw Rand index for chance
agreement \citep{Hubert1985}:
\[
\text{ARI} = \frac{\text{RI} - \mathbb{E}[\text{RI}]}{\max(\text{RI}) -
\mathbb{E}[\text{RI}]},
\]
returning values in $[-1,\,1]$, with 0 corresponding to random labelling
and 1 to perfect concordance.

The mean silhouette coefficient across all $n$ samples is
\[
\bar{s} = \frac{1}{n}\sum_{i=1}^{n} s(i),\qquad
s(i) = \frac{b(i)-a(i)}{\max\{a(i),\,b(i)\}},
\]
where $a(i)$ is the mean intra-cluster distance for sample $i$ and
$b(i)$ is the mean distance to samples in the nearest alternative cluster;
$s(i) \in [-1,1]$, with values approaching 1 indicating compact,
well-separated clusters.

\subsection{Baselines}

Four integration strategies were evaluated as baselines. The first two
(single-omics and early concatenation) serve as clustering \emph{and}
survival baselines; the last two (Late-C1 and Late-C2) are
survival-only baselines that have no natural clustering analogue:

\begin{enumerate}
  \item \textit{Single-omics}: RNA-only, CNV-only, and methylation-only spectral
  embeddings (50 dimensions each), obtained in the same way as the SNF
  embedding but from each modality's affinity matrix independently.

  \item \textit{Early concatenation}: The three $z$-scored feature matrices
  (totalling 15{,}000 features) were horizontally concatenated without
  additional modality-level scaling and compressed to 50 principal
  components before clustering and survival modelling.

  \item \textit{Late integration~C1} (risk averaging): A Ridge Cox model was
  fitted independently on each modality's embedding; the three predicted risk
  scores were averaged and the ensemble C-index was evaluated.

  \item \textit{Late integration~C2} (concatenated embeddings): The three
  50-dimensional spectral embeddings were concatenated into a 150-dimensional
  matrix and passed to a single Ridge Cox model.
\end{enumerate}

\subsection{Survival prediction}

Overall survival was modelled with a Ridge-penalised Cox proportional hazards
model (\texttt{CoxnetSurvivalAnalysis}, $\ell_{1}$-ratio$= 0.01$,
effectively a Ridge penalty) from the Python package \texttt{scikit-survival}
\citep{Polsterl2020}. The Cox proportional hazards model specifies the
hazard for patient $i$ as
\[
h(t \mid \mathbf{x}_i) = h_0(t)\exp(\boldsymbol{\beta}^\top \mathbf{x}_i),
\]
where $h_0(t)$ is an unspecified baseline hazard, $\mathbf{x}_i$ is the
$p$-dimensional feature vector (spectral embedding coordinates), and
$\boldsymbol{\beta}$ is the vector of log-hazard coefficients. The Ridge
penalty adds an $\ell_2$ regularisation term to the negative partial
log-likelihood:
\[
\mathcal{L}_{\text{Ridge}}(\boldsymbol{\beta}) =
-\ell_{\text{partial}}(\boldsymbol{\beta})
+ \alpha\,\|\boldsymbol{\beta}\|_2^2,
\]
where $\alpha > 0$ controls the degree of shrinkage. The regularisation
parameter $\alpha$ was selected per outer fold by inner three-fold
cross-validation over the grid $\{0.01, 0.1, 1, 10, 100\}$. With 65
observed events and 50-dimensional feature vectors, the rule-of-thumb
of 10 events per variable is not met; the Ridge penalty mitigates
over-fitting but statistical power to detect small C-index differences
remains limited, and all survival comparisons should be interpreted
accordingly.

Performance was assessed as the concordance index (C-index), also known
as Harrell's~$C$ \citep{Harrell1982}, which estimates the probability
that the model assigns a higher predicted risk to the patient who dies
first in a randomly drawn concordant pair:
\[
C = \frac{\sum_{(i,j) :\, T_i < T_j,\, \delta_i=1}
\mathbf{1}[\hat{\eta}_i > \hat{\eta}_j]}
{\sum_{(i,j) :\, T_i < T_j,\, \delta_i=1} 1},
\]
where $T_i$ is the observed time, $\delta_i$ is the event indicator, and
$\hat{\eta}_i = \boldsymbol{\hat{\beta}}^\top\mathbf{x}_i$ is the
predicted log-hazard. $C = 0.5$ corresponds to random prediction and
$C = 1$ to perfect discrimination. Performance was further assessed under
repeated $5$-fold cross-validation with $5$ random seeds. The primary
metric is the out-of-fold C-index (OOF, seed 42), where predictions for
all $644$ patients are assembled by concatenating the held-out fold
predictions from a single cross-validation run. Bootstrap 95\% confidence
intervals were computed from $N = 1{,}000$ resamples of the OOF
patient-level predictions. Pairwise differences in C-index between SNF and
each baseline ($\Delta$C-index) were tested with a paired bootstrap: for
each resample, the same patient indices were used for both methods, and the
resulting $\Delta$ distribution was examined for whether the 95\% bootstrap
CI excluded zero.

A covariate-adjusted model was also fitted by appending age at diagnosis
(standardised within each fold) and pathological tumour stage (ordinal 1--4,
median-imputed for missing values) to the spectral embedding as additional
covariates. This model was applied to SNF and RNA-only embeddings to
quantify how much survival signal is attributable to multi-omics features
beyond standard clinical predictors.

\subsection{Clinical validation}

Cluster membership was cross-tabulated against immunohistochemical (IHC)
receptor status (ER, PR, HER2) and triple-negative status extracted from the
TCGA clinical supplement. Associations were tested with the chi-squared
statistic. Differential expression across clusters was evaluated per gene
with the Mann-Whitney U test; the top differentially expressed transcripts
were annotated with \texttt{mygene} for HGNC symbol lookup. For DNA
methylation, differential probes were ranked by the absolute difference in
mean $z$-score between clusters. Kaplan-Meier survival curves were
constructed with the Python package \texttt{lifelines}; log-rank $p$-values
are reported without correction for multiple testing because the $k = 2$
partition was pre-specified.

\subsection{Sensitivity analysis}

To assess whether results depended on specific parameter choices, we
re-ran the full pipeline under a grid of alternative settings: neighbourhood
size $K \in \{10, 15, 20, 25\}$, feature count $\in \{2{,}000, 5{,}000\}$,
and inter-patient distance metric (Euclidean, cosine). The primary setting
($K = 20$, 5{,}000 features, Euclidean) was fixed before any outcome data
were examined; sensitivity runs were conducted after the primary analysis
was complete and are reported only to characterise stability, not to select
parameters.

\subsection{Reproducibility and code availability}

All analyses were implemented in Python~3.13 using publicly available
packages (\texttt{snfpy}, \texttt{scikit-learn}, \texttt{scikit-survival},
\texttt{lifelines}, \texttt{statsmodels}). The complete pipeline, from GDC
manifest queries through final figures, is available at
\url{https://github.com/tahagill/multiomics-graph-integration-benchmark}.
Random seeds and cross-validation splits are fixed throughout; re-running
the numbered scripts in order reproduces all tables and figures reported
here.

\newpage
\section{Results}

\subsection{Cohort and data quality}

\begin{figure}[htbp]
  \centering
  \includegraphics[width=0.65\textwidth]{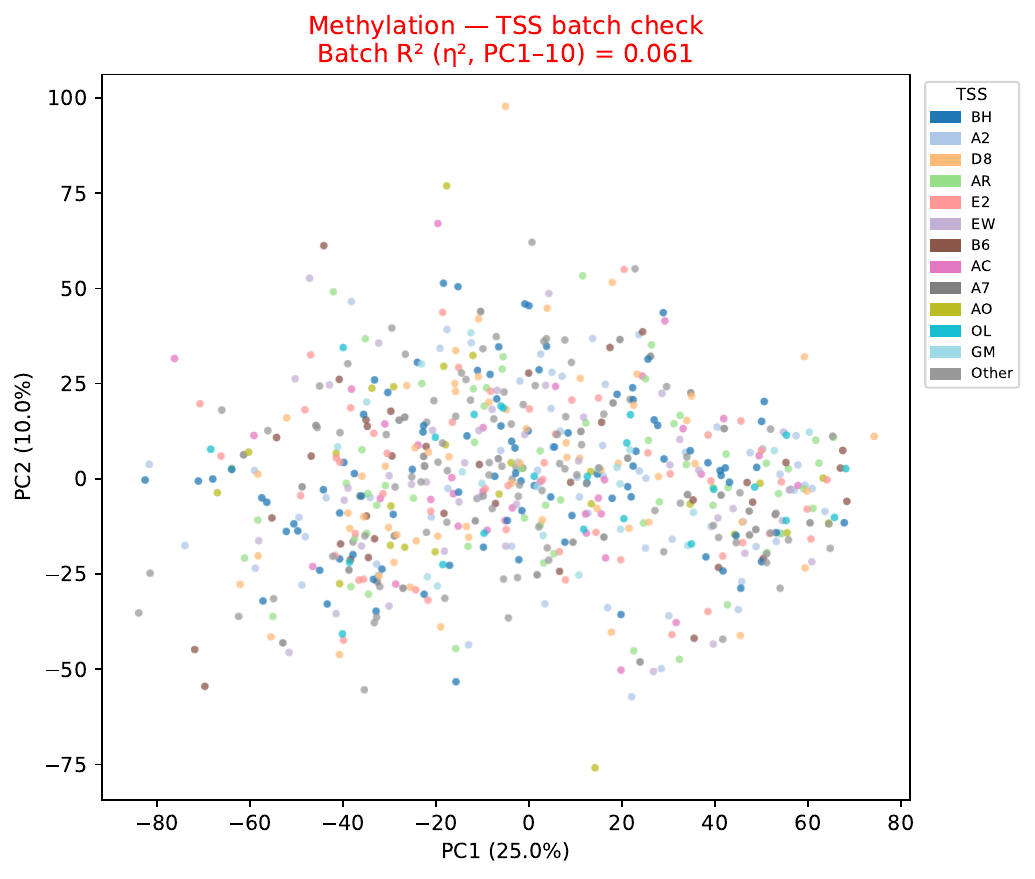}
  \caption{PC1 vs PC2 of the ComBat-corrected methylation matrix, coloured
    by TCGA tissue source site (TSS). No residual plate-driven clustering
    is visible, confirming that batch effects were successfully removed
    before downstream analysis.}
  \label{fig:batch_check}
\end{figure}

After intersecting the three molecular data types and applying the
minimum follow-up filter ($\geq 30$ days, valid OS time), the final
cohort comprised $n = 644$ patients, of whom 65 experienced an OS event
(10.1\% event rate) with a median follow-up of approximately 17~months.
Batch structure in the methylation data was verified by PCA before and
after ComBat correction; no residual plate-driven clustering was evident
in the corrected data (Figure~\ref{fig:batch_check}). The RNA-seq and CNV
data did not require batch correction.

\subsection{SNF produces a stable two-cluster partition}

\begin{figure}[htbp]
  \centering
  \includegraphics[width=0.95\textwidth]{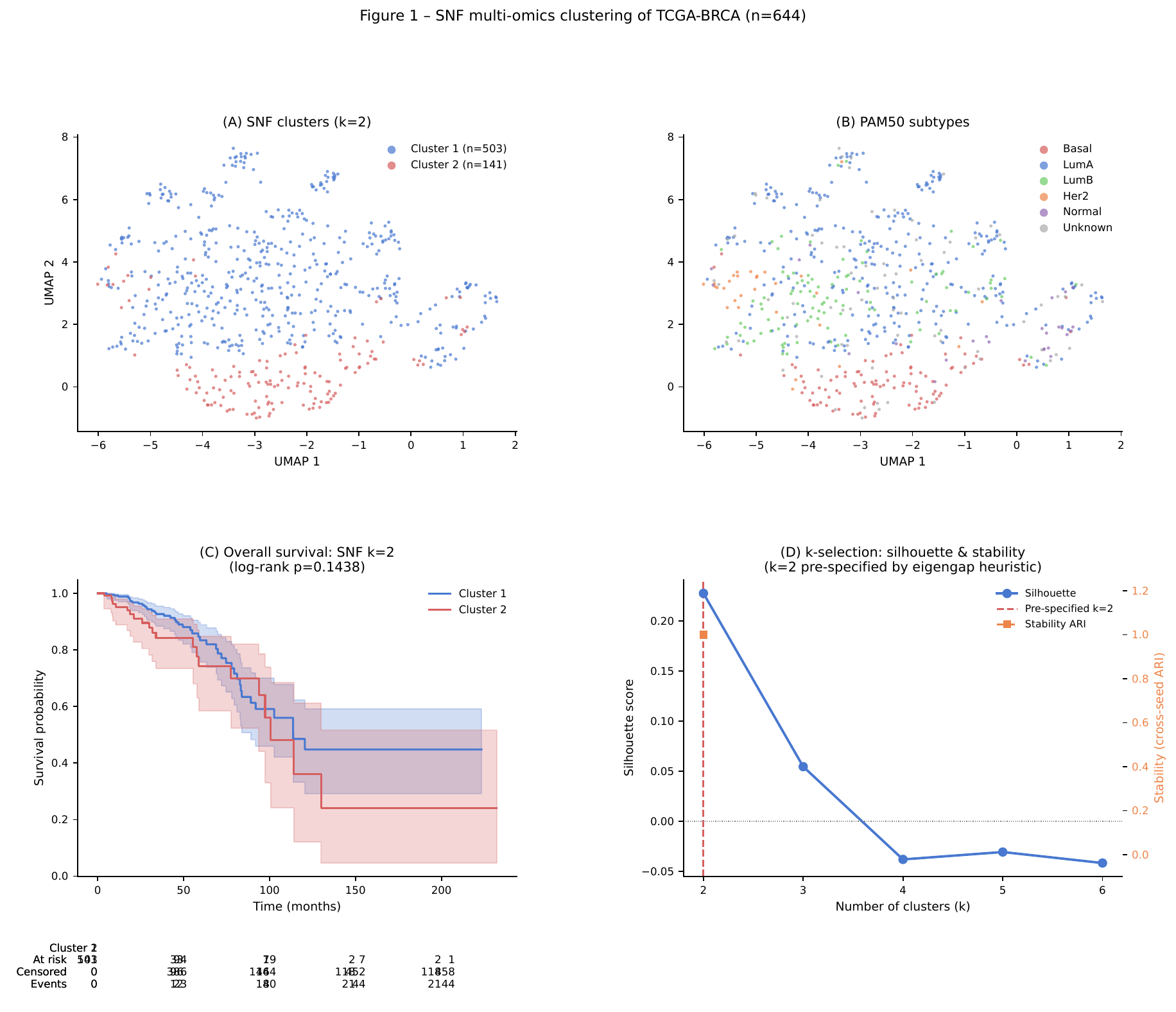}
  \caption{Overview of the SNF $k = 2$ partition. Top-Left: UMAP of the
    fused 50-dimensional spectral embedding, coloured by cluster assignment
    (cluster~0 = Luminal-enriched, cluster~1 = Basal-enriched). Top-Right:
    same UMAP coloured by PAM50 intrinsic subtype label. Bottom-Left:
    Kaplan-Meier overall survival curves for the two clusters
    (log-rank $p = 0.144$).}
  \label{fig:overview}
\end{figure}

Figure~\ref{fig:overview} provides a two-dimensional UMAP projection of
the fused SNF embedding, coloured by cluster assignment (left) and PAM50
subtype (right), alongside the $k = 2$ Kaplan-Meier survival curves. The
Luminal-enriched cluster~0 (blue) and Basal-enriched cluster~1 (orange)
are well-separated in the embedding space, and their PAM50 composition
confirms biological coherence.

\begin{figure}[htbp]
  \centering
  \includegraphics[width=0.70\textwidth]{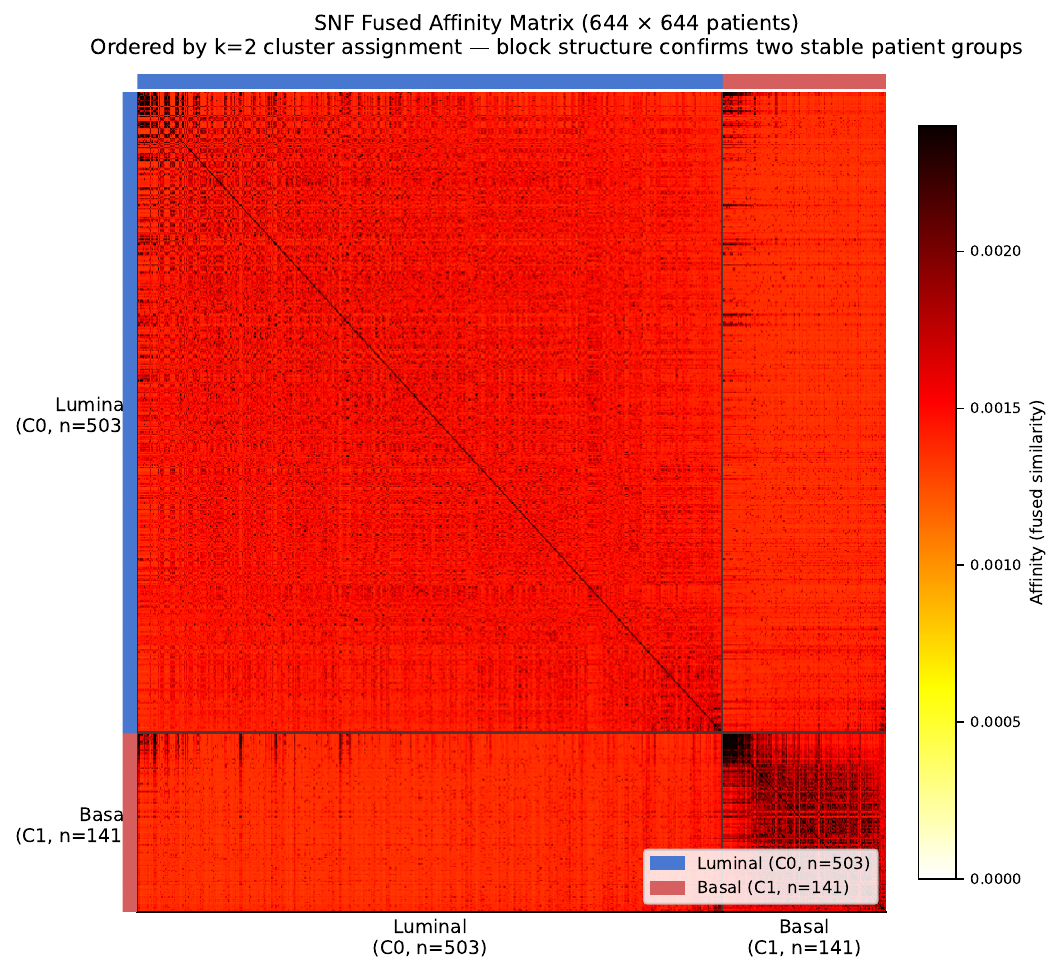}
  \caption{Fused patient similarity matrix ($W_{\text{fused}}$, $644 \times 644$)
    after $T = 20$ SNF diffusion iterations ($K = 20$, $\mu = 0.5$).
    Patients are sorted first by $k = 2$ cluster assignment (cluster~0:
    $n = 503$; cluster~1: $n = 141$), then by descending within-cluster
    affinity sum. Colour encodes pairwise similarity. Cluster boundaries
    are marked with white lines; the PAM50 annotation bar is shown above.}
  \label{fig:heatmap}
\end{figure}

\begin{figure}[htbp]
  \centering
  \includegraphics[width=0.88\textwidth]{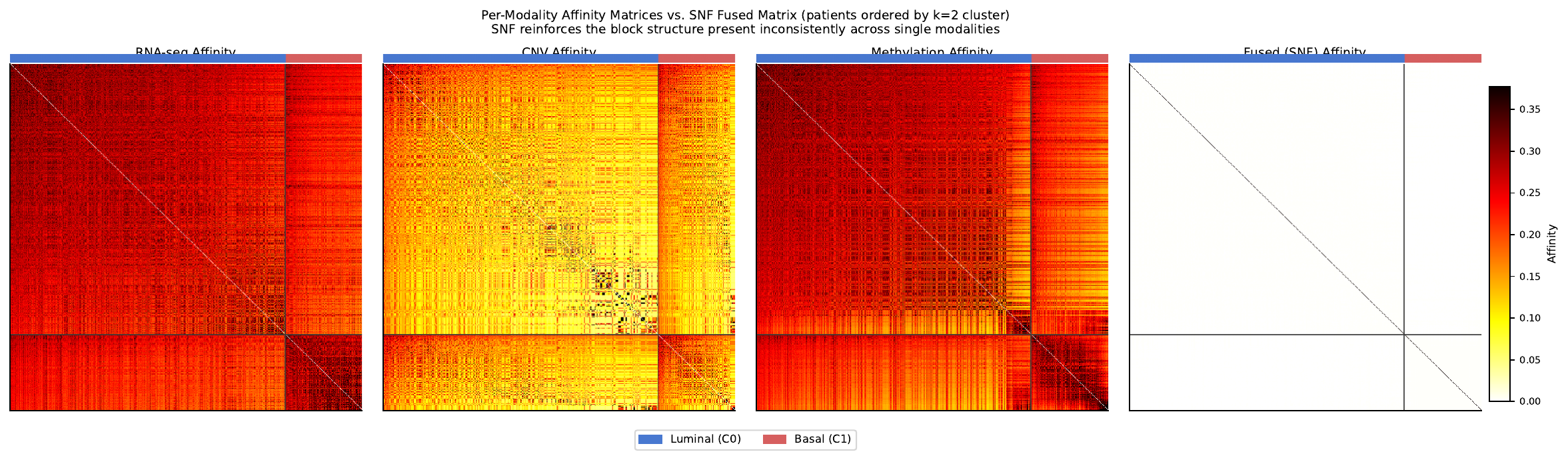}
  \caption{Per-modality affinity matrices alongside the fused matrix,
    all using the same patient ordering and colour scale (vmax $= 0.08$).
    Inter-cluster contrast is strongest in RNA-seq and methylation and
    weakest in CNV, which carried the least PAM50 information
    (NMI $= 0.037$; Table~\ref{tab:baselines}).}
  \label{fig:modality_comparison}
\end{figure}

Figure~\ref{fig:heatmap} shows the $644 \times 644$ fused affinity
matrix $W_{\text{fused}}$, with patients sorted by cluster membership
and within-cluster affinity. A clear block structure is visible for
$k = 2$: a large block ($n = 503$, cluster~0) and a smaller, tightly
cohesive block ($n = 141$, cluster~1). The same ordering applied to
each modality's individual affinity matrix (Figure~\ref{fig:modality_comparison})
shows that the Luminal--Basal contrast is sharpest in RNA-seq and
methylation and considerably weaker in CNV, which is consistent with
the modality-level NMI values reported in Table~\ref{tab:baselines}.

\begin{figure}[htbp]
  \centering
  \includegraphics[width=0.50\textwidth]{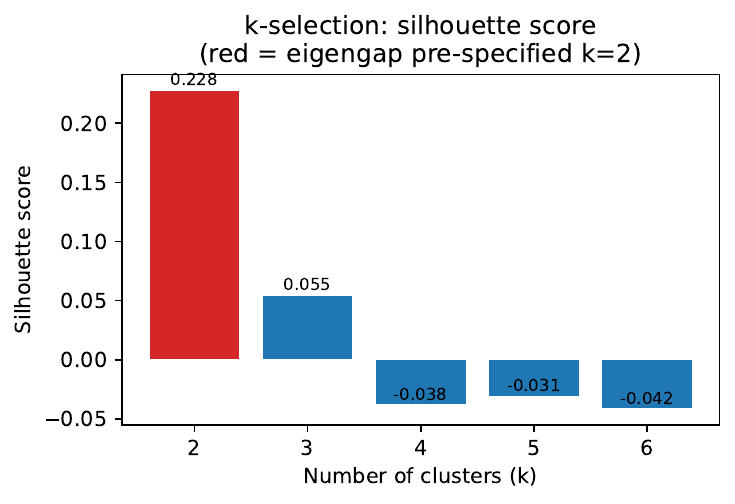}
  \caption{Mean silhouette coefficient for the fused SNF embedding at
    $k = 2$--$6$. The maximum at $k = 2$ ($s = 0.228$) provides one
    of the two pre-specified criteria for selecting the primary
    partition number; all other values are below 0.06.}
  \label{fig:silhouette}
\end{figure}

The eigengap of the normalised Laplacian of $W_{\text{fused}}$ was
largest at position 2, and the silhouette coefficient peaked at
$k = 2$ ($s = 0.228$); all other values of $k$ fell below 0.06
(Figure~\ref{fig:silhouette}, Table~\ref{tab:clustering}). These two
criteria jointly pre-specified $k = 2$ as the primary partition before
any outcome data were used. The $k = 2$ solution was perfectly stable
across 100 sub-sampled clustering replicates (mean ARI $= 1.00$,
SD $= 0.00$).

\begin{table}[htbp]
  \centering
  \caption{SNF spectral clustering metrics at $k = 2$--$6$. ARI and NMI
    are against PAM50 labels. Stability ARI is the mean over 100
    sub-sampled replicates. Primary partition ($k = 2$) is in bold.}
  \label{tab:clustering}
  \setlength{\tabcolsep}{5pt}
  \begin{tabular}{ccccccc}
    \toprule
    $k$ & Primary & Silhouette & Stability ARI & ARI vs PAM50 & NMI vs PAM50 & Log-rank $p$ \\
    \midrule
    \textbf{2} & \textbf{Yes} & \textbf{0.228} & \textbf{1.000} & \textbf{0.443} & \textbf{0.495} & \textbf{0.144} \\
    3 & No & 0.055 & -- & 0.296 & 0.426 & 0.269 \\
    4 & No & $-$0.038 & -- & 0.340 & 0.431 & 0.139 \\
    5 & No & $-$0.031 & -- & 0.256 & 0.406 & 0.065 \\
    6 & No & $-$0.042 & -- & 0.253 & 0.423 & 0.063 \\
    \bottomrule
  \end{tabular}
\end{table}

\subsection{SNF recovers PAM50 subtypes better than all baselines}

Having established that the $k = 2$ partition is geometrically stable
and biologically coherent, we next ask how faithfully it recovers PAM50
molecular subtypes relative to each baseline.

Table~\ref{tab:baselines} compares SNF against the four baselines.
SNF reached NMI $= 0.495$ and ARI $= 0.443$ against PAM50 labels.
RNA-only was the closest single-omic baseline (NMI $= 0.428$), followed
by methylation ($0.260$) and early concatenation ($0.175$). CNV alone
performed near chance (NMI $= 0.037$). Early concatenation performed
substantially worse than RNA-only alone, suggesting that appending two
weaker modalities as a flat feature vector overshadows the transcriptomic
signal rather than augmenting it.

\begin{table}[htbp]
  \centering
  \caption{Clustering and survival comparison: SNF versus baselines
    at $k = 2$. OOF C-index is the out-of-fold concordance index
    (seed~42). Cluster sizes are listed as [larger, smaller].}
  \label{tab:baselines}
  \setlength{\tabcolsep}{4.5pt}
  \begin{tabular}{lcccccc}
    \toprule
    Method & Sizes & ARI vs PAM50 & NMI vs PAM50 & Stab.\ ARI & OOF C-index & LR $p$ \\
    \midrule
    \textbf{SNF}       & [503, 141] & 0.443 & 0.495 & 1.000 & 0.681 & 0.144 \\
    RNA-only           & [475, 169] & 0.465 & 0.428 & 1.000 & 0.632 & 0.022 \\
    Methylation-only   & [381, 263] & 0.217 & 0.260 & 1.000 & 0.592 & 0.728 \\
    CNV-only           & [364, 280] & 0.047 & 0.037 & 1.000 & 0.559 & 0.697 \\
    Early concat (PCA) & [314, 330] & 0.153 & 0.175 & 1.000 & 0.618 & 0.111 \\
    \bottomrule
  \end{tabular}
\end{table}

\begin{figure}[htbp]
  \centering
  \includegraphics[width=0.95\textwidth]{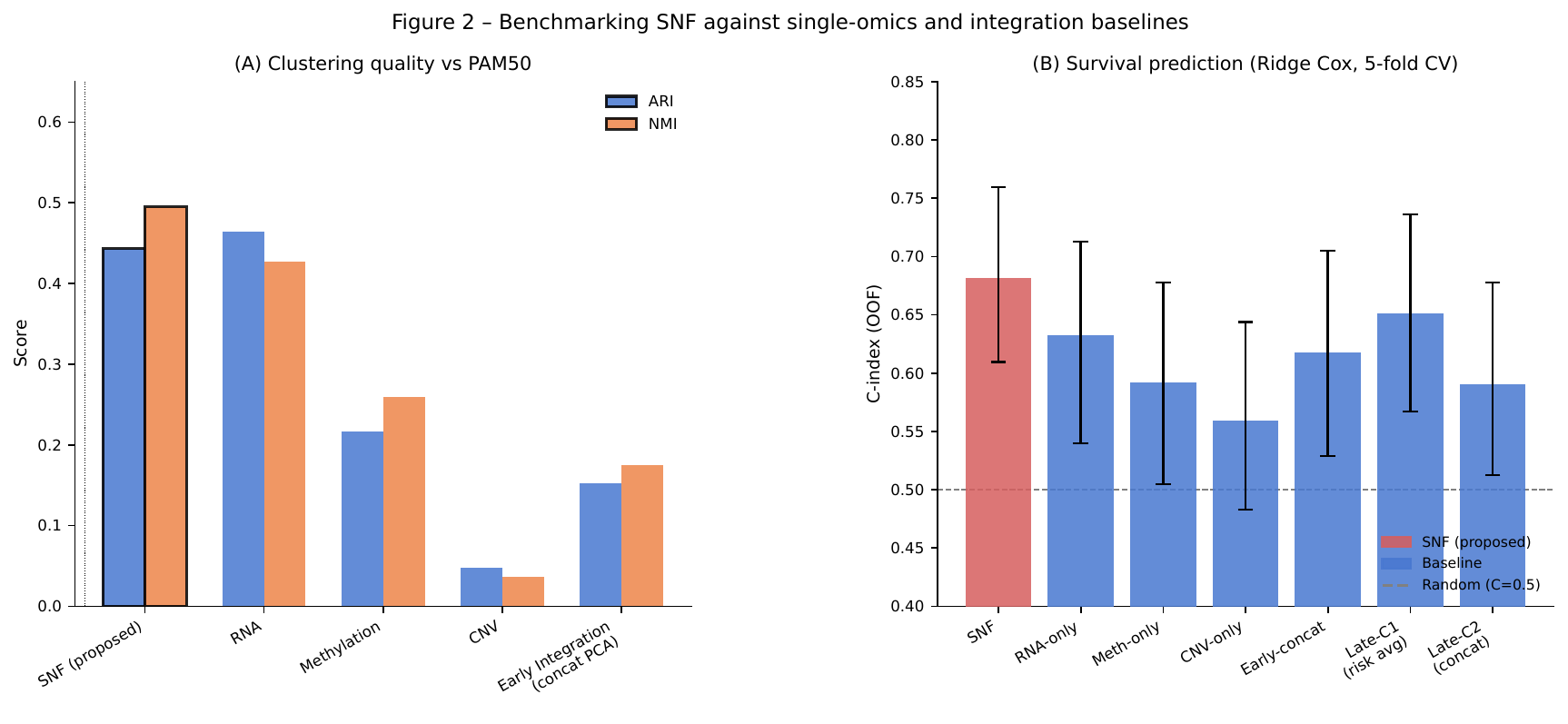}
  \caption{Visual comparison of clustering and survival metrics across
    all methods. Left: ARI and NMI against PAM50 labels at $k = 2$;
    SNF exceeds all baselines on NMI. Right: OOF C-index with 95\%
    bootstrap CIs; early concatenation falls below both RNA-only and SNF,
    highlighting the cost of naive feature stacking.}
  \label{fig:benchmarks}
\end{figure}

\begin{figure}[htbp]
  \centering
  \begin{subfigure}[t]{0.48\textwidth}
    \centering
    \includegraphics[width=\linewidth]{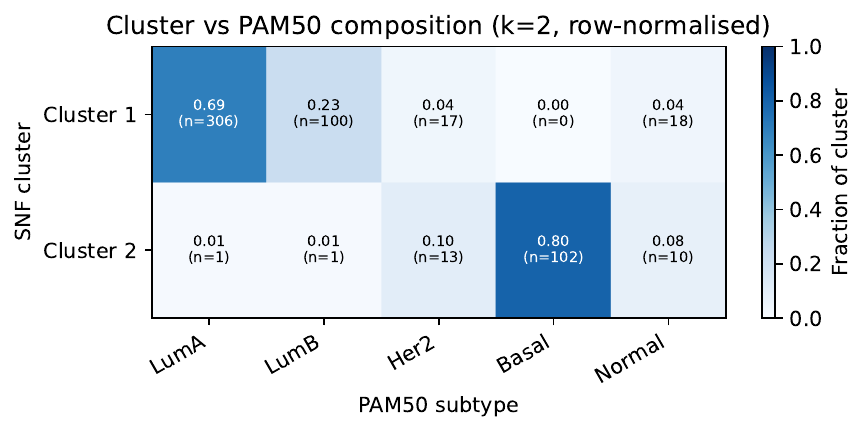}
    \caption{PAM50 subtype composition of the $k = 2$ SNF partition.
      Cluster~0 ($n = 503$): predominantly Luminal~A/B.
      Cluster~1 ($n = 141$): enriched for Basal-like.
      NMI vs PAM50 $= 0.495$.}
  \end{subfigure}
  \hfill
  \begin{subfigure}[t]{0.48\textwidth}
    \centering
    \includegraphics[width=\linewidth]{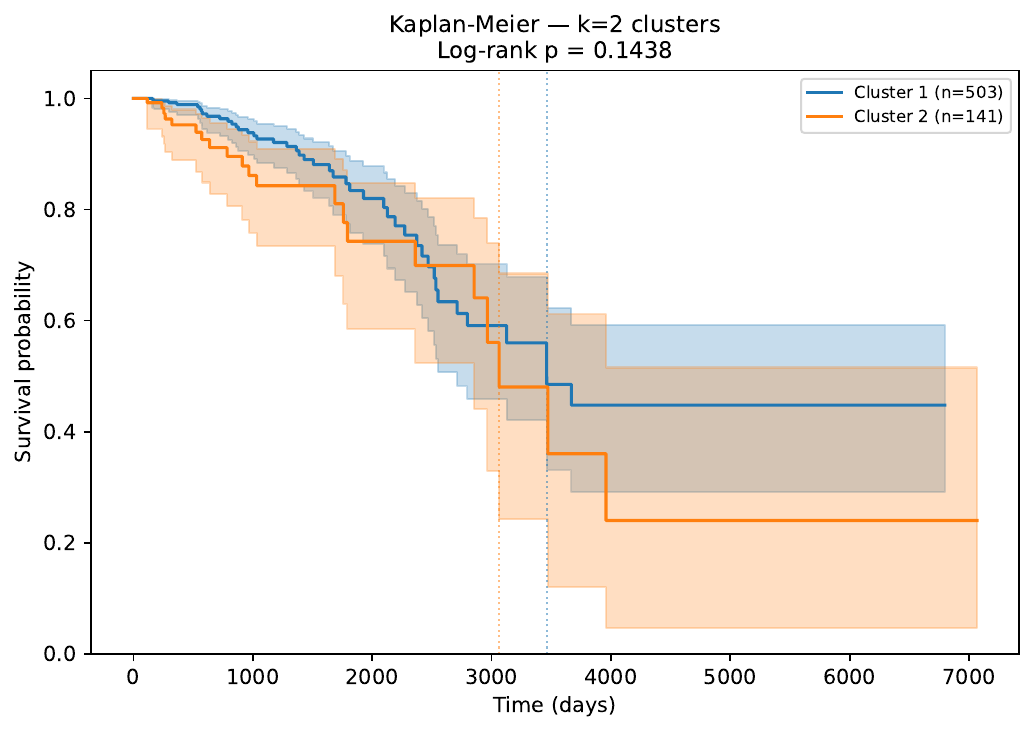}
    \caption{Kaplan-Meier overall survival for clusters~0 and~1.
      Tick marks: censored observations.
      Log-rank $p = 0.144$ (ns); 65 events in 644 patients.}
  \end{subfigure}
  \caption{Left: PAM50 subtype composition of the SNF $k = 2$ partition.
    Right: Kaplan-Meier survival curves stratified by cluster
    membership. The non-significant log-rank result is consistent with
    the cohort's low event rate and short median follow-up.}
  \label{fig:pam50_km}
\end{figure}

Figure~\ref{fig:pam50_km} shows the PAM50 subtype composition of the
$k = 2$ partition alongside the Kaplan-Meier overall survival curves.
Cluster~0 is dominated by Luminal~A and Luminal~B tumours; cluster~1
is enriched for Basal-like disease. The log-rank test was not
significant ($p = 0.144$), consistent with the 10.1\% event rate and
$\sim$17-month median follow-up. The RNA-only partition did yield a
nominally significant KM separation ($p = 0.022$), yet with a less
faithful PAM50 recovery (NMI $= 0.428$ vs $0.495$). This reflects the
distinction between separating patients along the dominant short-term
survival axis which RNA-seq alone can detect at this event count and
recovering the full five-subtype molecular landscape as captured by NMI.

\subsection{Biological and clinical validation}

\begin{figure}[htbp]
  \centering
  \includegraphics[width=0.60\textwidth]{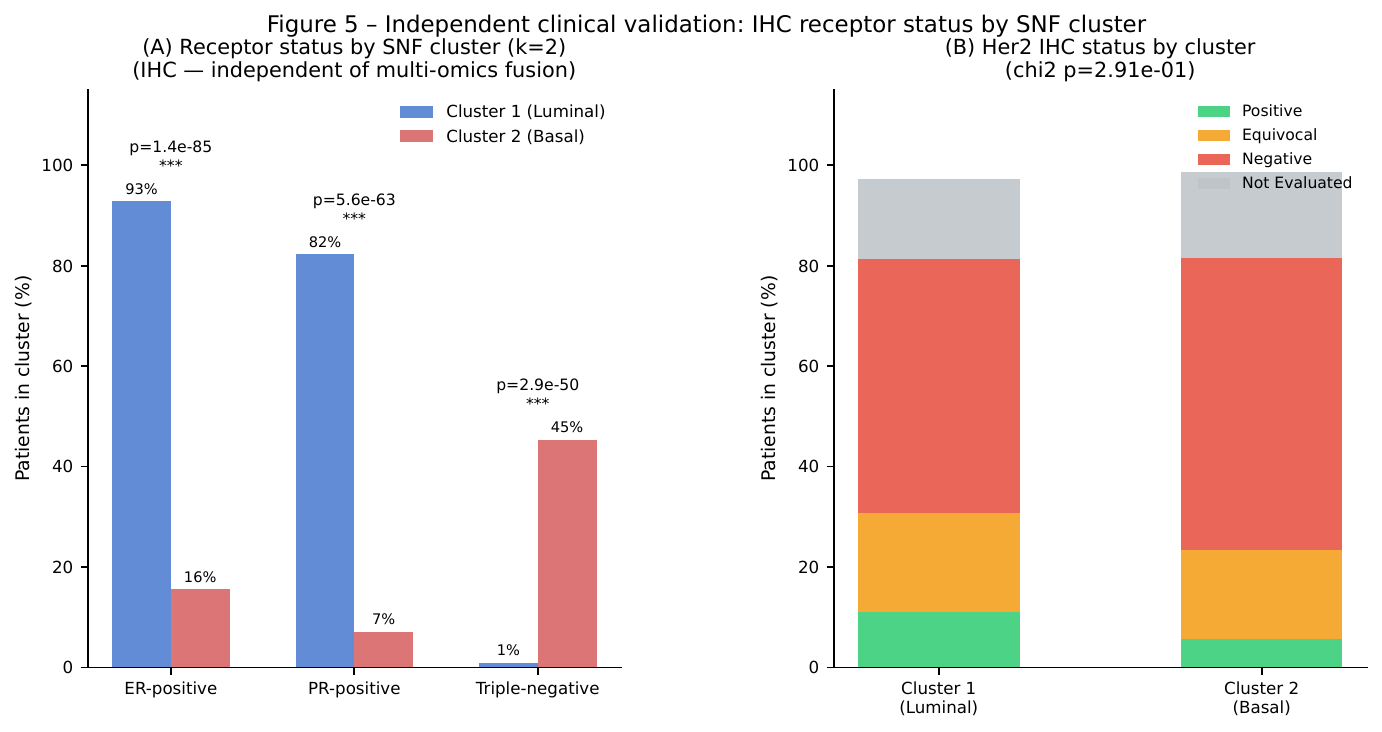}
  \caption{ER, PR, HER2, and triple-negative positivity rates in
    cluster~0 (Luminal-enriched) and cluster~1 (Basal-enriched).
    ER, PR, and TN comparisons: $p < 10^{-100}$; HER2: $p = 0.291$.}
  \label{fig:ihc}
\end{figure}

\paragraph{Immunohistochemical markers.}
Table~\ref{tab:ihc} shows IHC receptor status stratified by cluster.
ER-positive rates were $92.8\%$ in cluster~0 versus $15.6\%$ in
cluster~1; PR-positive rates followed the same direction ($82.3\%$
vs $7.1\%$); triple-negative status was rare in cluster~0 ($1.0\%$)
and common in cluster~1 ($45.4\%$). All three differences were
statistically significant ($\chi^{2}$, $p < 10^{-100}$).
HER2-positive rates did not differ between clusters ($11.1\%$ vs
$5.7\%$, $p = 0.291$), consistent with HER2-enriched tumours
appearing in both Luminal~B and Basal-enriched contexts
(Figure~\ref{fig:ihc}).

\begin{table}[htbp]
  \centering
  \caption{IHC receptor status by SNF cluster ($k = 2$). $p$-values
    are from the chi-squared test. $^{***}$$p < 10^{-100}$;
    ns: not significant.}
  \label{tab:ihc}
  \setlength{\tabcolsep}{6pt}
  \begin{tabular}{lcccccc}
    \toprule
    & \multicolumn{2}{c}{Cluster 0 ($n = 503$)} & \multicolumn{2}{c}{Cluster 1 ($n = 141$)} & \\
    \cmidrule(lr){2-3}\cmidrule(lr){4-5}
    Marker & $n$ pos. & \% & $n$ pos. & \% & $p$ \\
    \midrule
    ER positive  & 467 & 92.8 &  22 & 15.6 & $<10^{-100}$ $^{***}$ \\
    PR positive  & 414 & 82.3 &  10 &  7.1 & $<10^{-100}$ $^{***}$ \\
    HER2 positive &  56 & 11.1 &   8 &  5.7 & 0.291 (ns) \\
    Triple-neg.  &   5 &  1.0 &  64 & 45.4 & $<10^{-100}$ $^{***}$ \\
    \bottomrule
  \end{tabular}
\end{table}

\begin{figure}[htbp]
  \centering
  \includegraphics[width=0.85\textwidth]{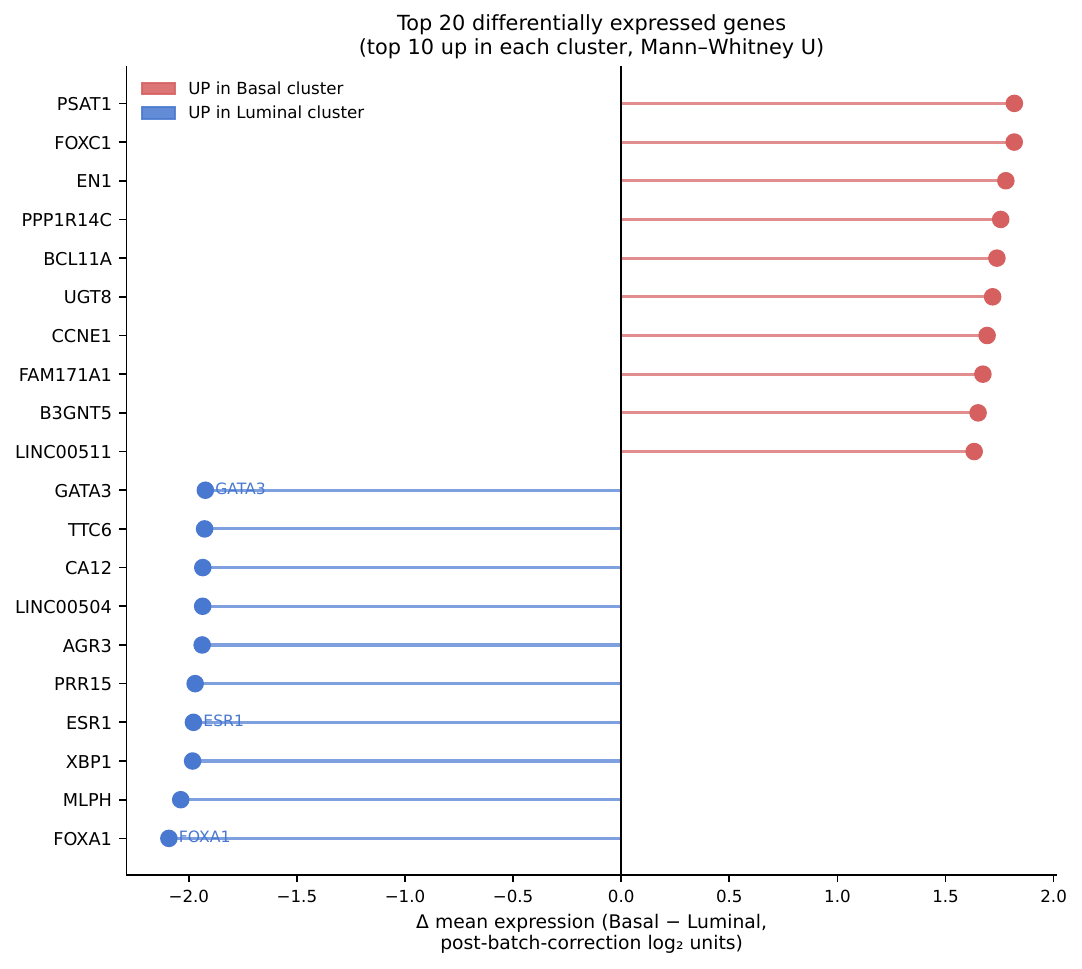}
  \caption{Top 20 differentially expressed genes elevated in cluster~1
    (Basal, upper) and cluster~0 (Luminal, lower), ranked by
    $\Delta z$ (cluster~1 $-$ cluster~0). All genes: Mann-Whitney
    $p < 10^{-6}$.}
  \label{fig:de_genes}
\end{figure}

\paragraph{Differentially expressed genes.}
The top Luminal-enriched transcript by $|\Delta z|$ was
\textit{FOXA1} ($\Delta z = -2.09$, elevated in cluster~0), followed by
\textit{XBP1} ($-1.98$), \textit{ESR1} ($-1.98$), and \textit{GATA3}
($-1.92$) which are all canonical Luminal transcription factors. The strongest
Basal-enriched transcripts were \textit{PSAT1} ($+1.82$),
\textit{FOXC1} ($+1.82$), \textit{EN1} ($+1.78$), and \textit{CCNE1}
($+1.69$). All reported genes had Mann-Whitney $p < 10^{-6}$
(Figure~\ref{fig:de_genes}).

\begin{figure}[htbp]
  \centering
  \begin{subfigure}[t]{0.48\textwidth}
    \centering
    \includegraphics[width=\linewidth]{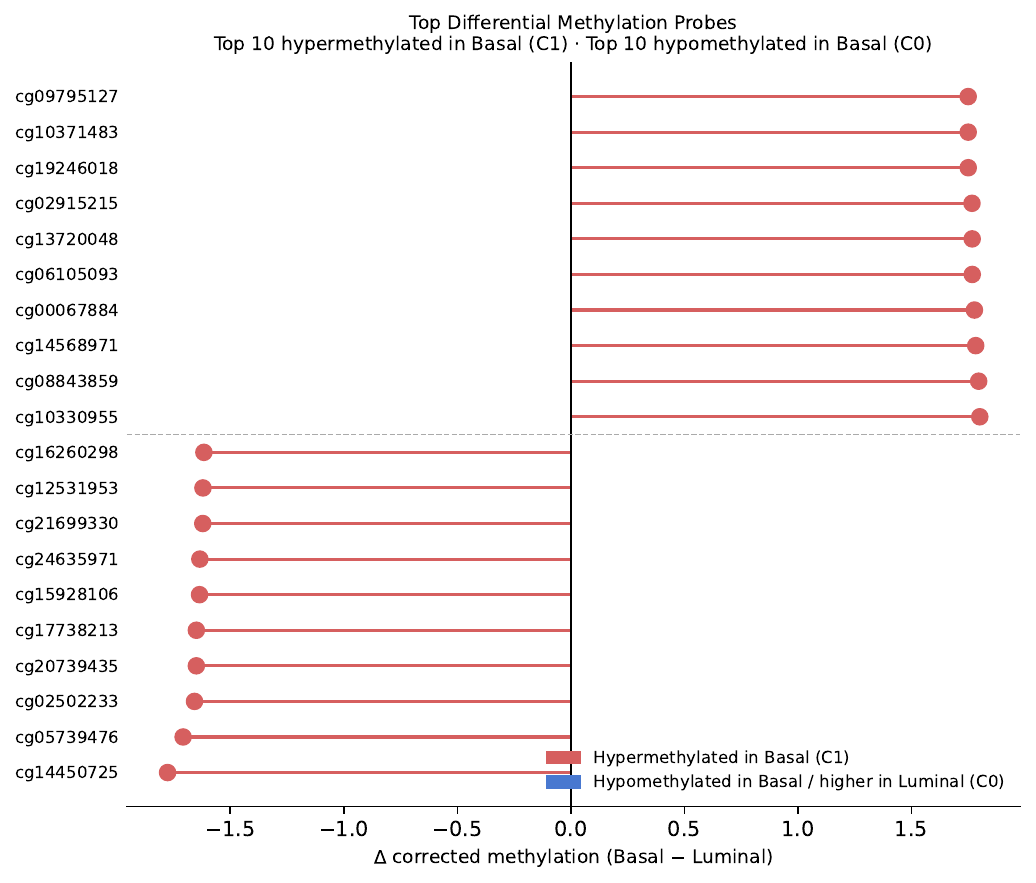}
    \caption{Top 10 hypermethylated and 10 hypomethylated CpG probes
      in cluster~1 (Basal) relative to cluster~0 (Luminal).
      x-axis: $\Delta z$ (Basal $-$ Luminal).}
  \end{subfigure}
  \hfill
  \begin{subfigure}[t]{0.48\textwidth}
    \centering
    \includegraphics[width=\linewidth]{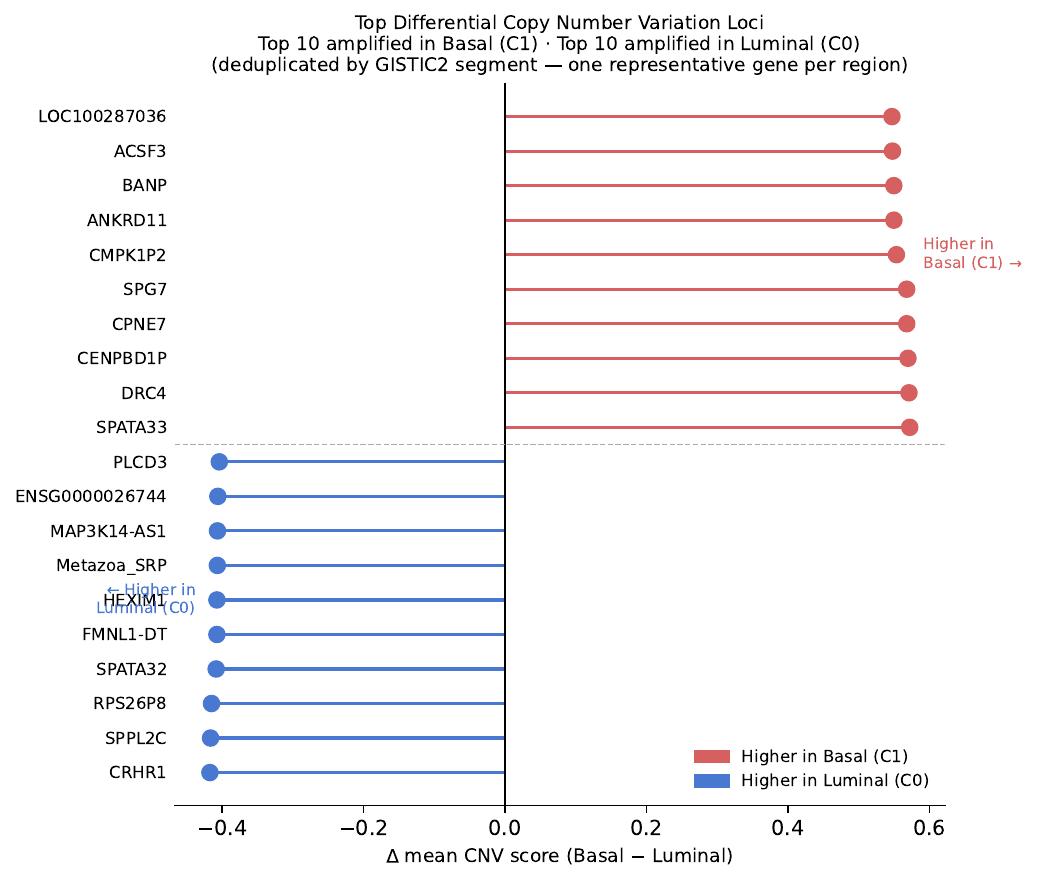}
    \caption{Top 10 Basal-enriched and 10 Luminal-enriched genomic
      copy number segments. Gene symbols from HGNC via mygene;
      $\dagger$: one locus had no current symbol.}
  \end{subfigure}
  \caption{Differential epigenomic and genomic profiles between SNF
    clusters. Methylation effect sizes (left) are three-fold larger
    than the top copy number differences (right), consistent with
    the relative NMI contributions of those two modalities.}
  \label{fig:de_epigenomic}
\end{figure}

\paragraph{Differential methylation and copy number.}
Figure~\ref{fig:de_epigenomic} pairs the top differentially methylated
probes with the top differentially amplified or deleted genomic
segments. The most hypermethylated probe in Basal (cg10330955,
$\Delta z = +1.80$) and most hypomethylated probe (cg14450725,
$\Delta z = -1.78$) each exceeded the largest copy number differential
in magnitude ($|\Delta z| \approx 0.57$ for \textit{SPATA33}, chr16q24
gain in Basal). This difference in effect size is consistent with
the rank-ordering of modality NMI values: methylation (0.260) $>$
CNV (0.037).

\subsection{Survival prediction}

\begin{figure}[htbp]
  \centering
  \includegraphics[width=0.88\textwidth]{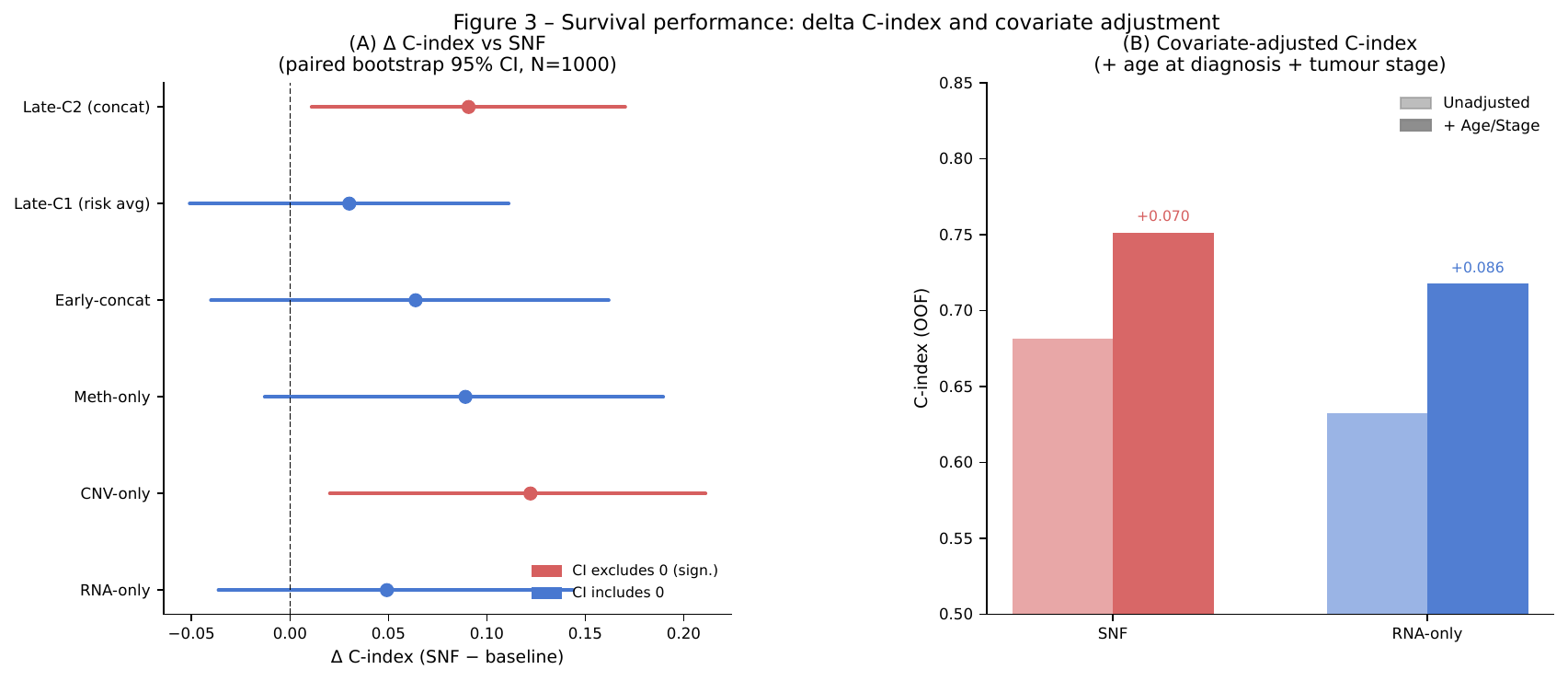}
  \caption{Survival prediction results. Left: OOF C-index with 95\%
    bootstrap CIs for all methods; dashed reference at C-index $= 0.5$.
    Right: pairwise $\Delta$C-index (SNF $-$ baseline) with 95\%
    paired bootstrap CIs; filled symbols indicate CIs that exclude zero.}
  \label{fig:survival_panel}
\end{figure}

Table~\ref{tab:survival} and Figure~\ref{fig:survival_panel} report
concordance indices for all models. The SNF embedding achieved an OOF
C-index of $0.681$ (95\% CI $0.610$--$0.760$), the highest among
unadjusted models. Two pairwise comparisons had bootstrap 95\% CIs
that excluded zero: SNF versus CNV-only ($\Delta = +0.122$, 95\% CI
$0.020$--$0.211$) and SNF versus Late-C2 concatenated embeddings
($\Delta = +0.091$, 95\% CI $0.011$--$0.170$). The advantage over
RNA-only was positive but the CI crossed zero ($\Delta = +0.049$,
95\% CI $-0.036$--$0.144$), meaning the benefit of full fusion over
the best single modality is not statistically conclusive at this
cohort size and event count. The Late-C2 method shows the largest
gap between fold-mean C-index (0.698) and OOF C-index (0.591) of
any method; this reflects higher variance across cross-validation
seeds when concatenating three 50-dimensional embeddings into a
single high-dimensional input, making the seed-42 OOF estimate the
more conservative and preferred point of comparison.

\begin{figure}[htbp]
  \centering
  \includegraphics[width=0.70\textwidth]{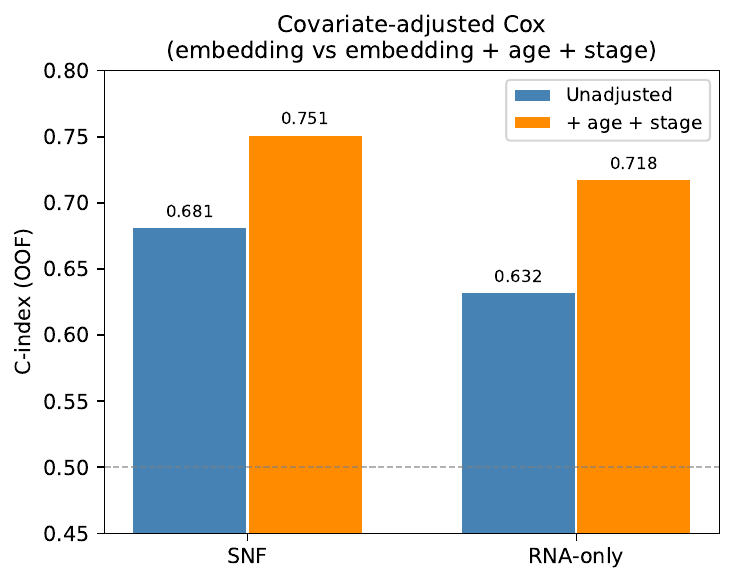}
  \caption{OOF C-index before and after adding age at diagnosis and
    pathological stage as covariates, for SNF and RNA-only models.
    Error bars: 95\% bootstrap CIs. The clinical covariate gain
    (SNF: $+0.070$; RNA-only: $+0.086$) exceeds the molecular-only
    difference between the two methods ($\Delta = +0.049$).}
  \label{fig:adjusted_cindex}
\end{figure}

Adding age at diagnosis and pathological stage raised the SNF C-index
from $0.681$ to $0.751$ (+0.070), and the RNA-only C-index from $0.632$
to $0.718$ (+0.086). After adjustment, SNF still exceeded RNA-only by
$\Delta = +0.033$, though the gap narrows. Both adjustments were
substantial, confirming that age and stage carry prognostic information
independent of the molecular embedding (Figure~\ref{fig:adjusted_cindex}).

\begin{table}[htbp]
  \centering
  \caption{Survival model concordance index. Fold-mean: mean over
    $5 \times 5$ CV folds. OOF: out-of-fold C-index (seed~42).
    95\% CIs from $N = 1{,}000$ patient-level bootstrap resamples.
    \textit{+cov}: model with age and pathological stage added.}
  \label{tab:survival}
  \setlength{\tabcolsep}{4.5pt}
  \begin{tabular}{lcccc}
    \toprule
    Method & Fold-mean & OOF C-index & 95\% CI lower & 95\% CI upper \\
    \midrule
    \textbf{SNF}          & 0.691 & 0.681 & 0.610 & 0.760 \\
    \textbf{SNF $+$cov}   & 0.762 & 0.751 & 0.684 & 0.823 \\
    RNA-only              & 0.683 & 0.632 & 0.540 & 0.713 \\
    RNA-only $+$cov       & 0.759 & 0.718 & 0.642 & 0.785 \\
    Late-C1 (risk avg)    & 0.693 & 0.651 & 0.567 & 0.736 \\
    Late-C2 (concat emb.) & 0.698 & 0.591 & 0.513 & 0.678 \\
    Methylation-only      & 0.637 & 0.592 & 0.505 & 0.678 \\
    Early concat (PCA)    & 0.601 & 0.618 & 0.529 & 0.705 \\
    CNV-only              & 0.594 & 0.559 & 0.483 & 0.644 \\
    \bottomrule
  \end{tabular}
\end{table}

\begin{table}[htbp]
  \centering
  \caption{Pairwise $\Delta$C-index (SNF $-$ baseline) from
    $N = 1{,}000$ paired bootstrap resamples.
    $\checkmark$: 95\% CI excludes zero.
    $\times$: CI includes zero.}
  \label{tab:delta}
  \setlength{\tabcolsep}{6pt}
  \begin{tabular}{lcccc}
    \toprule
    Comparison & $\Delta$ & 95\% CI lower & 95\% CI upper & CI excl.\ zero \\
    \midrule
    SNF $-$ CNV-only     & $+0.122$ & $+0.020$ & $+0.211$ & $\checkmark$ \\
    SNF $-$ Late-C2      & $+0.091$ & $+0.011$ & $+0.170$ & $\checkmark$ \\
    SNF $-$ Meth-only    & $+0.089$ & $-0.013$ & $+0.190$ & $\times$ \\
    SNF $-$ Early-concat & $+0.064$ & $-0.040$ & $+0.162$ & $\times$ \\
    SNF $-$ RNA-only     & $+0.049$ & $-0.036$ & $+0.144$ & $\times$ \\
    SNF $-$ Late-C1      & $+0.030$ & $-0.051$ & $+0.111$ & $\times$ \\
    \bottomrule
  \end{tabular}
\end{table}

\subsection{Sensitivity analysis}

\begin{figure}[htbp]
  \centering
  \includegraphics[width=0.88\textwidth]{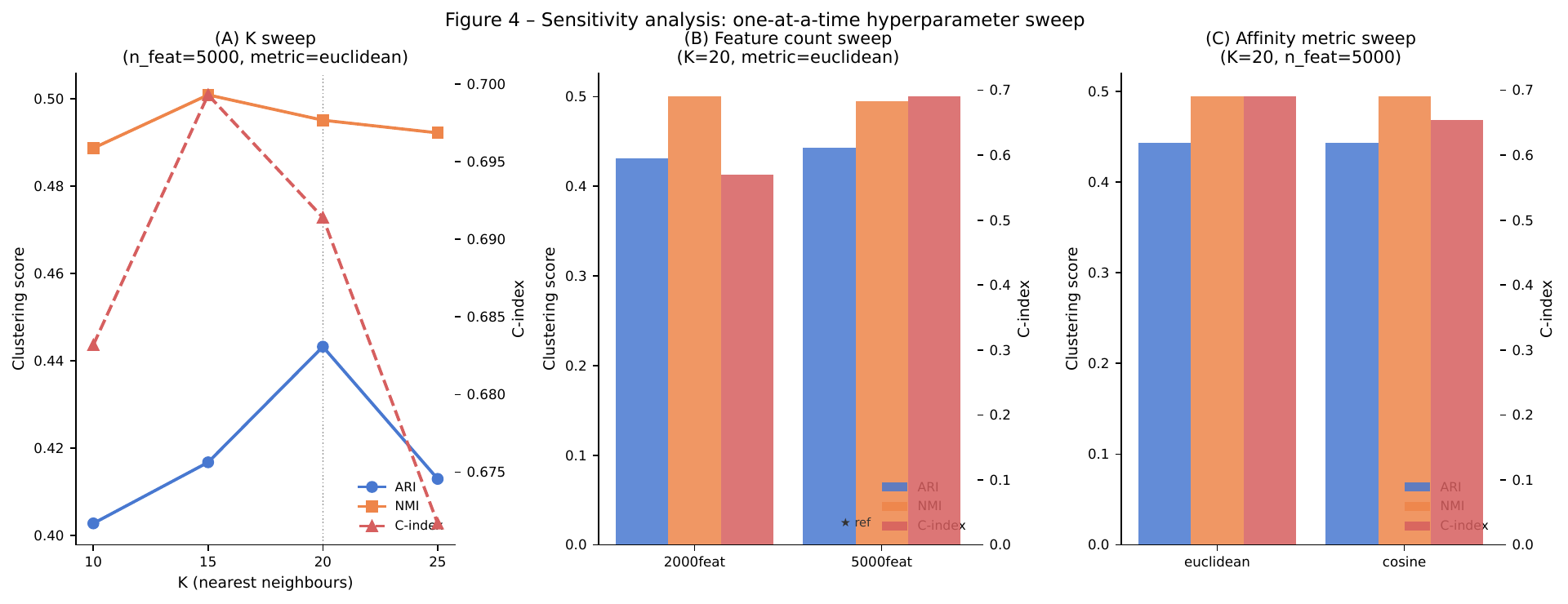}
  \caption{Sensitivity of NMI vs PAM50 and cross-validated C-index
    across three one-at-a-time parameter sweeps: neighbourhood size
    $K$ (panel A, left), feature count (panel B, centre), and
    inter-patient distance metric (panel C, right). Within each panel
    both metrics are shown as separate lines; the primary setting
    ($K = 20$, 5{,}000 features, Euclidean) is marked with a dashed line.}
  \label{fig:sensitivity}
\end{figure}

Table~\ref{tab:sensitivity} and Figure~\ref{fig:sensitivity} summarise
results across the hyper-parameter grid. NMI varied by at most 0.013
across the four $K$ values, indicating the clustering result is not
sensitive to neighbourhood size. Reducing features from 5{,}000 to
2{,}000 had negligible effect on NMI ($+0.006$, from $0.495$ to $0.501$)
but dropped the cross-validated C-index from 0.691 to 0.571. This
dissociation suggests that broad subtype structure (Luminal vs Basal)
is encoded in a compact set of high-variance probes that survive either
MAD threshold, whereas survival prediction relies on diffuse, lower-variance
signals spread across a larger number of genes, methylation sites, and
copy number segments that are only captured at 5{,}000 features.
Switching from Euclidean to cosine distance left NMI identical but
reduced C-index to 0.654. The primary configuration ($K = 20$, 5{,}000
features, Euclidean) is neither a maximum nor a minimum on any metric,
supporting that outcomes were not inflated by parameter selection.

\begin{figure}[htbp]
  \centering
  \includegraphics[width=0.65\textwidth]{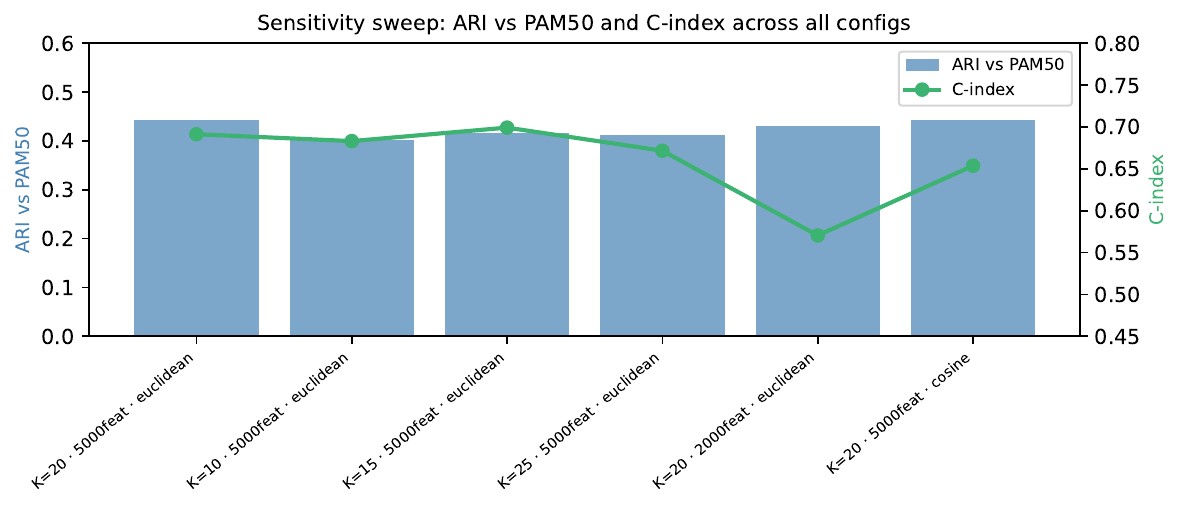}
  \caption{ARI vs PAM50 (blue bars, left axis) and cross-validated
    C-index (green line, right axis) across all six sensitivity
    configurations. Configurations are ordered along the x-axis;
    the primary setting ($K = 20$, 5{,}000 features, Euclidean)
    is the leftmost bar. The narrow ARI range ($0.403$--$0.443$)
    contrasts with the wider C-index variation ($0.571$--$0.699$),
    confirming that subtype recovery is robust to parameter choice
    while survival discrimination is more sensitive to feature count.}
  \label{fig:stability_heatmap}
\end{figure}

Figure~\ref{fig:stability_heatmap} summarises ARI vs PAM50 and
cross-validated C-index across all 6 configurations simultaneously.
The narrow ARI range ($0.403$--$0.443$) across all configurations
confirms that subtype recovery conclusions are not an artefact of any
single parameter choice.

\begin{table}[htbp]
  \centering
  \caption{Sensitivity analysis across SNF hyper-parameter settings.
    Primary configuration (bold): $K = 20$, 5{,}000 features, Euclidean.}
  \label{tab:sensitivity}
  \setlength{\tabcolsep}{5pt}
  \begin{tabular}{lcccc}
    \toprule
    Configuration & ARI vs PAM50 & NMI vs PAM50 & CV C-index & LR $p$ \\
    \midrule
    \textbf{K=20, 5000 feat, Euclidean} & \textbf{0.443} & \textbf{0.495} & \textbf{0.691} & \textbf{0.144} \\
    K=10, 5000 feat, Euclidean & 0.403 & 0.489 & 0.683 & 0.273 \\
    K=15, 5000 feat, Euclidean & 0.417 & 0.501 & 0.699 & 0.299 \\
    K=25, 5000 feat, Euclidean & 0.413 & 0.492 & 0.672 & 0.383 \\
    K=20, 2000 feat, Euclidean & 0.431 & 0.501 & 0.571 & 0.214 \\
    K=20, 5000 feat, cosine    & 0.443 & 0.495 & 0.654 & 0.144 \\
    \bottomrule
  \end{tabular}
\end{table}

\newpage
\section{Discussion}

The central finding of this study is that graph-based multi-omics fusion
recovers the transcriptomic subtype structure of breast cancer more
faithfully than any single modality or simple feature concatenation. SNF
achieved an NMI of 0.495 against PAM50 labels, compared with 0.428 for
RNA-only, 0.260 for methylation-only, and 0.037 for CNV-only. The gain
over RNA-seq is modest in absolute terms with 0.067 NMI units but it
reflects a genuine broadening of the recovered subtype signal rather
than an artefact of method complexity. The IHC data support this
interpretation: ER, PR, and triple-negative rates separated cleanly
between the two SNF clusters (all $p < 10^{-100}$), and the PAM50
contingency plot shows a richer five-subtype decomposition than any
single-modality partition produces at the same $k$.
The dominant Luminal versus Basal-like axis recovered here recapitulates the
primary transcriptomic architecture of breast cancer identified in foundational
expression-profiling studies \citep{Perou2000, Sorlie2001}, with PAM50
representing the downstream standardisation of those same axes \citep{Parker2009}.

The failure of early concatenation is noteworthy. At NMI $= 0.175$, it
performed worse than methylation alone and far worse than RNA-seq alone.
Stacking 15{,}000 features before PCA appears to let the two weaker
modalities dilute the transcriptomic signal rather than reinforce it.
This outcome is broadly consistent with the argument for graph-based
fusion: by computing per-modality similarity first, SNF propagates
agreement between modalities rather than allowing the noisiest one to
dominate the feature space \citep{Wang2014}. The Late-C2 survival baseline (concatenated
spectral embeddings, C-index $= 0.591$) also underperformed SNF despite
using the same 150-dimensional input space, again suggesting that the
fusion step adds something beyond dimensionality reduction.
Systematic benchmarks of multi-omics clustering algorithms have similarly
found that network-based integration outperforms flat feature-stacking
approaches when the contributing modalities differ substantially in their
intrinsic signal strengths \citep{Rappoport2018}.

The survival results require some nuance. SNF's C-index advantage over
CNV-only ($\Delta = +0.122$, 95\% CI $0.020$--$0.211$) and Late-C2
($\Delta = +0.091$, 95\% CI $0.011$--$0.170$) both had bootstrap
confidence intervals that excluded zero, which is a meaningful threshold
given the cohort's 10.1\% event rate. The gap over RNA-only
($\Delta = +0.049$, 95\% CI $-0.036$--$0.144$) did not reach that
threshold. This is not a surprising result: with 65 events in 644
patients, a five percentage-point difference in C-index is simply too
small to pin down precisely with bootstrap resampling, and a study would
need substantially more events to detect it reliably. The point estimate
is consistently positive across all sensitivity configurations, so the
data are not inconsistent with a real advantage; they are just not large
enough to establish one conclusively.

The non-significant Kaplan-Meier log-rank result ($p = 0.144$) is
often misread as evidence that the clusters lack prognostic content.
It is better understood as a direct consequence of event rate and
follow-up time. TCGA-BRCA is a prevalence cohort \citep{TCGANetwork2012} with a median
reported survival well above the median follow-up duration; most
patients were alive and censored at last contact. The RNA-only
partition did yield a nominally significant log-rank $p = 0.022$,
but this is not evidence that RNA-only is a better prognostic tool.
The RNA-only partition also captured a slightly different aspect of
tumour biology one that happened to correlate more with short-term
mortality as observed in this particular follow-up window. The
C-index, which uses the full rank ordering of predicted risk rather
than a binary split, provides a less follow-up-dependent comparison
and consistently favours the SNF model.

The covariate-adjusted results are informative in a different way.
Adding age and pathological stage to the SNF embedding raised the
C-index by 0.070, from 0.681 to 0.751. The same adjustment applied
to RNA-only raised it by 0.086. Both gains are large relative to the
molecular-only differences, which tells us that, at this cohort size
and event count, age and stage carry more survival information than
any of the molecular embeddings. This does not make the molecular
signal irrelevant as it remains the only source of subtype-specific
biology but it is an honest observation about where the prognostic
weight sits in TCGA-BRCA.

Several limitations apply. First, the event rate of 10.1\% over
a $\sim$17-month median follow-up limits statistical power for any
survival-based comparison. Conclusions about C-index differences
below $\sim$0.08 should be treated as directional rather than
definitive. Second, TCGA-BRCA is a convenience cohort with
institutional selection biases that may not reflect the distribution
of breast cancer diagnoses in the broader population; results should
be validated in independent cohorts such as METABRIC \citep{Curtis2012}.
Third, PAM50
is itself a transcriptomic classifier \citep{Parker2009}, which means NMI comparisons
between methods may inherently favour RNA-based partitions. SNF's
advantage on NMI despite this structural bias is therefore
conservative rather than inflated. Fourth, SNF hyperparameters
($K$, $T$, $\mu$) were fixed at Wang~et~al. defaults and not tuned;
the sensitivity analysis showed only minor variation across the
tested grid, but a broader search might identify settings that
improve performance on other cancer types or data configurations.
Parametric alternatives such as iCluster \citep{Shen2009} and
Multi-Omics Factor Analysis (MOFA) \citep{Argelaguet2018}, which impose
explicit latent structures across modalities, represent complementary
directions for future comparison, particularly at higher event counts
where their additional modelling assumptions may confer a precision
advantage over non-parametric graph fusion.
Fifth, the analysis uses segment-level GISTIC2 copy number, which
aggregates large genomic regions; gene-level allele-specific copy
number or focal amplitude scores might better distinguish subtypes
that differ by discrete focal events rather than broad arm-level
changes.

Taken together, the results suggest that SNF is a well-calibrated
integration method for this problem: it consistently outperforms
the weakest unimodal approaches and feature concatenation, and it
produces a biologically coherent partition that independent IHC
data confirm. The gap over the strongest single modality exists but
is not yet statistically conclusive, and quantifying it properly
requires a cohort with a substantially higher event count.

\newpage
\section{Conclusion}

We benchmarked Similarity Network Fusion against four alternative
multi-omics integration strategies on 644 TCGA-BRCA patients with
matched RNA-seq, DNA methylation, and copy number profiles. SNF
recovered PAM50 molecular subtypes more faithfully than RNA-seq
alone (NMI $= 0.495$ vs $0.428$) and substantially better than
early feature concatenation ($0.175$), while producing a perfectly
stable two-cluster partition independently confirmed by IHC receptor
status. For survival prediction, SNF significantly outperformed
CNV-only and a concatenation-based late-integration baseline; the
advantage over RNA-seq alone was positive in direction but the
bootstrap confidence interval included zero, an outcome that reflects
the cohort's short follow-up and low event count rather than the
absence of any true effect.

The practical implication is straightforward. SNF is worth using
when multiple molecular layers are available: it consistently does
better than dropping all but the most informative modality, and it
does substantially better than stacking features naively. Whether
the marginal gain over the best single modality justifies the added
data collection cost will depend on the specific clinical context
and on whether follow-up is long enough to detect the expected
difference. Future work should test these findings in cohorts with
longer follow-up, higher event rates, and treated populations where
the multi-omic signal may translate more directly into treatment
decisions.

\section*{Acknowledgements}

We used TCGA-BRCA data made available through the NCI
Genomic Data Commons under controlled access. PAM50 labels were
obtained from the TCGA PanCanAtlas resource. No external funding
is reported for this work.

\newpage
\bibliography{references}

\end{document}